\documentclass[fleqn,10pt]{wlscirep}
\usepackage[utf8]{inputenc}
\usepackage[T1]{fontenc}
\usepackage{lineno}

\usepackage{graphicx}
\usepackage{caption}

\title{Reprogrammable plasmonic topological insulators with ultrafast control}

\author[1,2,3]{Jian Wei You}
\author[2,3]{Qian Ma}
\author[1]{Zhihao Lan}
\author[2]{Qiang Xiao}
\author[1*]{Nicolae C. Panoiu}
\author[2*]{Tie Jun Cui}

\affil[1]{Department of Electronic and Electrical Engineering, University College London, \newline Torrington Place, London WC1E 7JE, United Kingdom.}

\affil[2]{State Key Laboratory of Millimeter Waves and Institute of Electromagnetic Space, Southeast University, \newline Nanjing 210096, China.}

\affil[3]{These authors contributed equally: Jian Wei You, Qian Ma}

\affil[*]{Email: n.panoiu@ucl.ac.uk, tjcui@seu.edu.cn}

\begin{abstract}
\textbf{Topological photonics has revolutionized our understanding of light propagation, providing a remarkably robust way to manipulate light. Despite the intensive research and rapid progress in this field, most of existing studies are focused on designing a static photonic structure to realize a specific topological functionality or phenomenon. Developing a dynamic and universal photonic topological platform to intelligently switch multiple topological functionalities at ultrafast speed is still a great challenge. Here we theoretically propose and experimentally demonstrate an ultrafast reprogrammable plasmonic topological insulator, where the topological propagation route can be dynamically changed at nanosecond-level switching time, which is more than $1\times10^7$ times faster than the current state-of-the-art, leading to an experimental demonstration of unprecedentedly ultrafast multi-channel optical analog-digital converter. This orders-of-magnitude improvement compared to previous works is due to the innovative use of ultrafast electric switches to implement the programmability of our plasmonic topological insulator, which enables us to precisely encode each unit cell by dynamically controlling its digital plasmonic states while keeping its geometry and material parameters unchanged. Our reprogrammable topological plasmonic platform can be fabricated by the widely-used printed circuit board technology, making it much more attractive and compatible with current highly integrated photoelectric systems. Furthermore, due to its flexible programmability, many existing photonic topological functionalities can be integrated into this versatile topological platform. Our work brings the current studies of photonic topological insulators to a digital and intelligent era, which could open new avenues towards the development of software-defined photoelectric elements in high-speed communications and computation-based intelligent devices with built-in topological protection.}
\end{abstract}
\begin{document}

\flushbottom
\maketitle
\thispagestyle{empty}
\captionsetup[figure]{labelfont={bf},labelformat={default},labelsep=period,name={Fig. }}

\section*{Introduction}

The discovery of topological states of matter in condensed matter physics has inspired the search for analogous effects in classical and bosonic systems, especially fruitful in photonics. The beginnings of topological photonics can be traced to the efforts to emulate the quantum Hall effect (QHE), where the time-reversal symmetry of the system is broken by exploiting magneto-optic photonic crystals under external static magnetic fields \cite{hk10RMP,ljs14NatPho,oza19RMP,hr08PRL,raha09PRA,wcjs09Nat}. However, magneto-optical effects in most magnetic materials are weak and in many cases, it can be cumbersome and inconvenient to implement a magnetic device. To overcome these limitations, nonmagnetic photonic topological systems are developed subsequently, in which the time-reversal symmetry is preserved, to mimic the quantum spin-Hall effect (QSHE) or quantum valley-Hall effect (QVHE) \cite{kha13NatMat,whx15PRL,nix18SA,hex19NatCom,swt19NatNano}. The unique topological features of light in topological photonics, such as unidirectional and backscattering-immune propagation against defects and sharp bends, offer a robust way to control the behavior of light. The study of topological photonics has revolutionized our understanding of light propagation, and holds great promise for practical photonic applications, especially for the development of robust linear and nonlinear photonic devices \cite{hmfm13NatPho,bah17Sci,mgh18Nat,ylp20SA,slck20APR}.

Currently, the studies of photonic topological insulators (PTIs) are mainly focused on the development of different static photonic systems to realize a specific photonic topological phenomenon or functionality \cite{yan19Nat,kru19NatNano,zen20Nat}, and hence their controllability and reconfigurability are limited. In practical applications, however, multiple photonic topological functionalities are expected to be achieved in a single but reconfigurable PTI, so that the time and costs associated with the design and fabrication process can be reduced. To this end, several key works on reconfigurable topological insulators \cite{che16NatMat,zha19Sci,cao19SB,xia18AdvMat,dcl18PNAS,zczh19EML} have been reported recently, in which the reconfigurability is mostly implemented by changing either the geometry or material parameters. Unfortunately, these features in PTIs cannot be easily altered once the photonic structure has been fabricated. Moreover, the mechanical, thermal or optical method used to realize the reconfigurability is not convenient for use in compact photoelectric integrated systems, and the reconfiguring speed is limited. Recently, the emergence of programmable metasurfaces opens a digital way to dynamically control the metasurface functionality, which has spurred wide interest in metamaterial community \cite{cqwz14LSA,gio14NatMat,lil17NatCom,zhan18NatCom,maq19LSA,maq20Pho}. However, limited by the reflection-type binary state, current programmable metasurfaces are mainly used to tailor the reflection and radiation of electromagnetic (EM) waves, and the study that aims to control the propagation route of guiding waves has been hardly reported.

\begin{figure}[b!]
\centering
\includegraphics[width=14.5 cm]{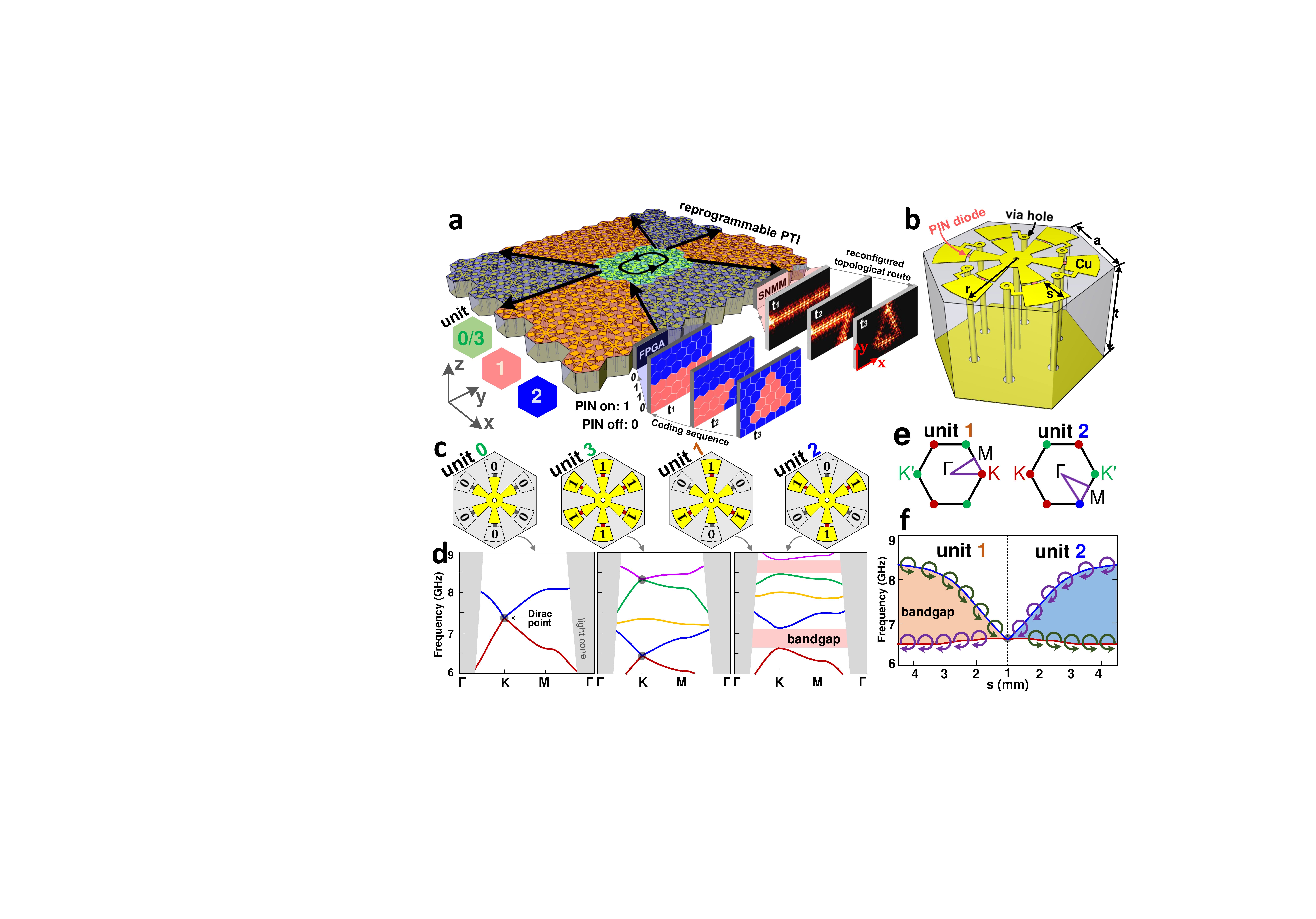}
\caption{\textbf{Design principle of RPTI.} \textbf{a}, Schematic of an ultrafast reprogrammable plasmonic topological insulator. The digital state of each reprogrammable unit cell is controlled electrically by the FPGA controller, which enables one to dynamically reconfigure the topological propagation route as needed. \textbf{b}, Reprogrammable unit cell. A six-arm metallic element and a ground plate perforated by six via holes are printed on the two facets of a dielectric spacer.  The binary states of the metallic element are controlled by six PIN diodes (shown in red). \textbf{c}, Four digital states of a designed 2-bit reprogrammable unit cell, whose odd (even) outer arms hold the same binary state.  \textbf{d}, Band diagrams of a crystal with the designed 2-bit unit cell. \textbf{e}, First Brillouin zone of digital unit-1 and unit-2. \textbf{f}, Topological phase transition and the valley-chirality property of the first and second bands of digital unit-1 and unit-2. The valley bandgap is enlarged when increasing the length $s$ of outer arm.}
\label{fig:PhysSche}
\end{figure}

In this work, taking advantage of both the flexible programmability of digital metasurfaces and the robust EM propagation of PTIs, we propose and experimentally demonstrate an ultrafast reprogrammable plasmonic topological insulator (RPTI), which is comprised of metallic elements arranged in a hexagonal lattice structure. To achieve an electrically driven programmability, each metallic element has six arms whose EM response can be electrically controlled via a positive-intrinsic-negative (PIN) diode. By switching the binary state of the PIN diode, a topological band gap can be created by breaking the space-inversion symmetry of the unit cell to emulate the QVHE. Moreover, each unit cell of the RPTI can be encoded independently to enable a dynamic control of topological light propagation routes. Compared to existing controllable PTIs, our RPTI has two unique features. First, its programmability is achieved by electrically encoding the binary states of the PIN diode, leading to an ultrafast dynamic control of different topological propagation routes. Second, our RPTI is fabricated by a widely-used printed circuit board (PCB) technology, and thus it can be seamlessly integrated with the commonly used PCB-based photoelectric integrated circuits. These unique features are crucial for the development of versatile and intelligent topological photoelectric devices for future practical applications.

\section*{The system}

The proposed RPTI (Fig. \ref{fig:PhysSche}a) consists of programmable unit cells (Fig. \ref{fig:PhysSche}b) arranged in a hexagonal lattice, with each unit cell being in one of its four digital states “0”, “1”, “2” and “3” (see Fig. \ref{fig:PhysSche}c). This system configuration enables us to encode one RPTI into several distinct plasmonic crystal domains, as shown in Fig. \ref{fig:PhysSche}a. Applying different coding sequences to each programmable unit cell at different times ($t_1$, $t_2$ and $t_3$) using a field-programmable gate array (FPGA), we can dynamically reconfigure the size and pattern of each plasmonic crystal domain to realize an on-demand manipulation of the domain-wall interface along which the microwave propagates. Moreover, this reconfigurable topological propagation route can be observed directly in experiments using a scanning near-field microwave microscopy (SNMM). Specifically, the programmable unit cell in Fig. \ref{fig:PhysSche}b consists of a dielectric spacer sandwiched between a controllable printed metallic element at the top and, at the bottom, a metallic ground plate perforated by six holes forming a hexagon. Unlike previous studies \cite{scui14LPR,wux17NatCom}, our disk has six inner arms and six outer arms, where each inner arm is bridged with one outer arm via a PIN diode. Furthermore, to ensure that the direct current (DC) bias voltage of the inner arms is zero, the six inner arms are connected to the ground plate by a metallic via hole. The six outer arms are connected to a FPGA control network by six metallic via holes. As a result, applying different DC voltages to each outer arm, one can dynamically switch “on” and “off” the corresponding PIN diode to emulate the binary states “1” and “0”, respectively. The geometry parameters of our RPTI studied in this work are $a=15/\sqrt{3}$ mm, $t=3$ mm and $r=7.5$ mm (see Fig. \ref{fig:PhysSche}b). In addition, the radius of the inner metallic circular disk is $1.5$ mm, and the gap distance between the inner and outer arms is $0.3$ mm.

In principle, the reprogrammable unit cell in Fig. \ref{fig:PhysSche}b is a 6-bit digital element possessing $2^6=64$ digital states. To simply our discussion, the odd (even) outer arms are designed to be in the same binary state as detailed in Extended Data Fig. 1, so that the 6-bit digital element can be reduced to a 2-bit digital element possessing $2^2=4$ digital states. In the context of programmable metasurfaces \cite{maq20Pho}, they could be naturally encoded as 0=“00”, 1=“01”, 2=“10” and 3=“11”, as listed in Fig. \ref{fig:PhysSche}c. Here, the binary codes “1” and “0” mean that the PIN diode is switched “on” and “off”, respectively. In the following, we denote each digital state as a digital unit in Fig. \ref{fig:PhysSche}c, though in practice they are the same programmable unit cell. Our RPTI is designed in the microwave regime, where metallic loss can be suppressed and spoof surface plasmons (SSP) exist \cite{pmg04Sci}. As demonstrated in Extended Data Figs. 2-5, the field of SSP is concentrated at the outer arm, if the corresponding PIN diode is switched on. However, if the diode is switched off, the field of SSP is located at the inner arm, and the corresponding outer arm is “invisible” to EM waves \cite{zhan20NSR}. Based on this feature, we can engineer and dynamical control the band diagram of the RPTI, as demonstrated in Fig. \ref{fig:PhysSche}d. In particular, due to the $C_6$ symmetry of the unit cell, there is a Dirac point in the band diagram of digital unit-0, in which all diodes are switched off. Similarly, Dirac points also appear in the band diagram of digital unit-3, in which all diodes are switch on. For the digital unit-1, the diodes on its odd (even) arms are switched off (on), so the $C_6$ symmetry is reduced to $C_3$ symmetry. As a result, the $C_6$-symmetry-protected Dirac points are gapped out to form bandgaps at $K$ and $K'$ valley points, as illustrated in the band diagram of digital unit-1. If the coupling between the $K$ and $K'$ valleys is neglected, the first band surrounding $K$ and $K'$ valleys is associated with opposite valley Chern numbers $\pm1/2$ \cite{wux17NatCom,qwu06PRB,eza13PRB}.

In the real space, the digital unit-1 and unit-2 are related by a $\pi$/3 in-plane rotation, thus their band diagrams are identical. However, in the wave-vector space, the first Brillouin zone (FBZ) of digital unit-2 can be obtained from that of digital unit-1 via a $\pi$/3 rotation (Fig. \ref{fig:PhysSche}e). As a result, the $K$($K'$) valley points of digital unit-2 correspond to the $K'$($K$) valley points of digital unit-1, and thus the difference of the valley Chern number across a domain-wall interface between a unit-1-type domain and a unit-2-type domain is $\pm1$. According to the principle of bulk-edge correspondence \cite{qwu06PRB,eza13PRB}, a nontrivial domain-wall interface mode emerges in the first bandgap. In Fig. \ref{fig:PhysSche}f, we further show that this nontrivial valley bandgap can be widened by increasing the length $s$ of the outer arm. Moreover, the valley-chirality property of the first and second bands of digital unit-1 is opposite to that of digital unit-2, with a topological phase transition occurring at $s=1$ mm.

\section*{Projected bands of domain-wall interfaces}

\begin{figure}[t!]
\centering
\includegraphics[width=15 cm]{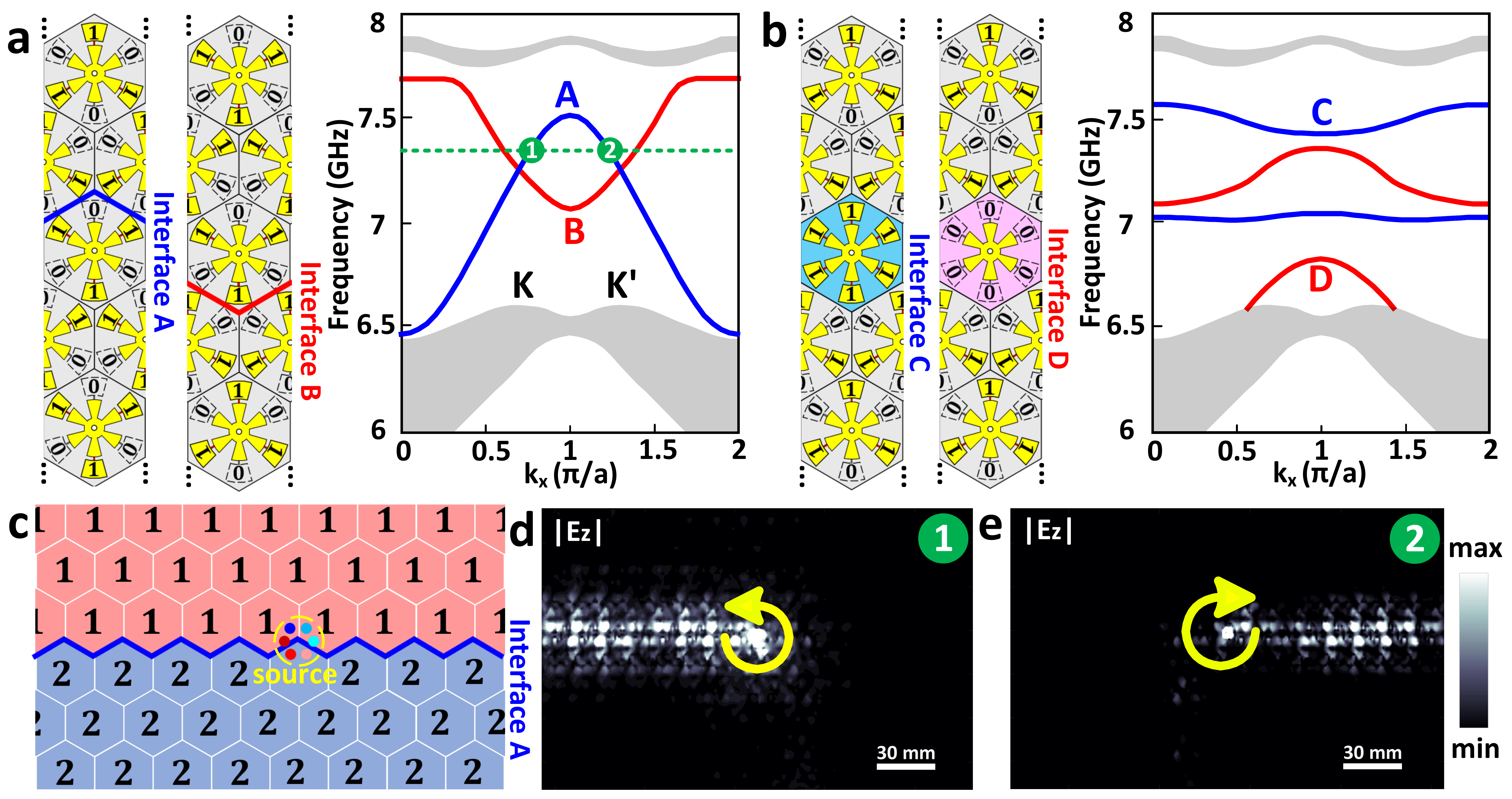}
\caption{\textbf{Domain-wall interface modes of RPTI and the unidirectional topological propagation feature.} \textbf{a}, Two topological interfaces and their projected band diagrams. The interfaces A and B separate two crystal domains, which are filled with digital unit-1 and unit-2, respectively. \textbf{b}, Two non-topological interfaces and their projected band diagrams. The interface C (D) is constructed by inserting a unit-3-type (unit-0-type) line defect into a photonic crystal comprised of digital unit-1. \textbf{c}, Straight waveguide with an A-type non-trivial interface, where a chiral source is employed. The symbols “1” and “2” denote digital unit-1 and unit-2, respectively. \textbf{d,e}, Simulated unidirectional propagations with the RCP (\textbf{d}) and LCP (\textbf{e}) excitations.} 
\label{fig:TopoInterf}
\end{figure}

To demonstrate the topological nature of nontrivial domain-wall interface modes within the first bandgap, we encode the unit cells of RPTI to construct two valley-Hall domain-wall interfaces with zigzag shape, namely the interfaces A and B in Fig. \ref{fig:TopoInterf}a, which separate two domains filled with digital unit-1 and unit-2, respectively. Here we use $s=2.5$ mm and the corresponding projected band diagram is given in the right panel of Fig. \ref{fig:TopoInterf}a. For each of the interfaces A and B, there is a valley-Hall interface mode in the first bandgap, where the blue (red) solid line indicates the nontrivial interface mode of the interface A (B), and the grey-shaded regions represent the projected bulk modes. In Fig. \ref{fig:TopoInterf}b, we also show two non-topological domain-wall interfaces in our RPTI, denoted by C and D, which are constructed by inserting a unit-3-type (for C) or unit-0-type (for D) line defect into a photonic crystal comprising digital unit-1 cells. We should note that, unlike previously investigated non-topological domain-wall interfaces, the route of our trivial domain-wall interface is dynamically reconfigurable. The corresponding projected band diagrams of interfaces C and D are given in the right panel of Fig. \ref{fig:TopoInterf}b, which shows that each non-topological domain-wall interface has two trivial interface-mode bands.

To study the property of unidirectional propagation of topological interface modes, a straight RPTI waveguide with the A-type interface is encoded in Fig. \ref{fig:TopoInterf}c, where a chiral source consisting of six electric dipoles with $\pm\pi/3$ phase difference between adjacent dipoles is employed to excite a right-circularly polarized (RCP) or left-circularly polarized (LCP) wave. As shown in Fig. \ref{fig:TopoInterf}a, if the source has frequency of 7.45 GHz, two distinct waveguide modes with opposite group-velocity, corresponding to points 1 and 2, are generated. Furthermore, from the valley-chirality-locking property of the topological interface modes, the chirality of the interface mode at point 1 is also opposite to that at point 2. Consequently, a unidirectional propagation of microwave can be achieved by selectively exciting an RCP or LCP phase vortex at the source point in Fig. \ref{fig:TopoInterf}c, which is numerically demonstrated in Figs. \ref{fig:TopoInterf}d and \ref{fig:TopoInterf}e.

\section*{Reconfigurable topological and non-topological light propagations}

\begin{figure}[t!]
\centering
\includegraphics[width=15 cm]{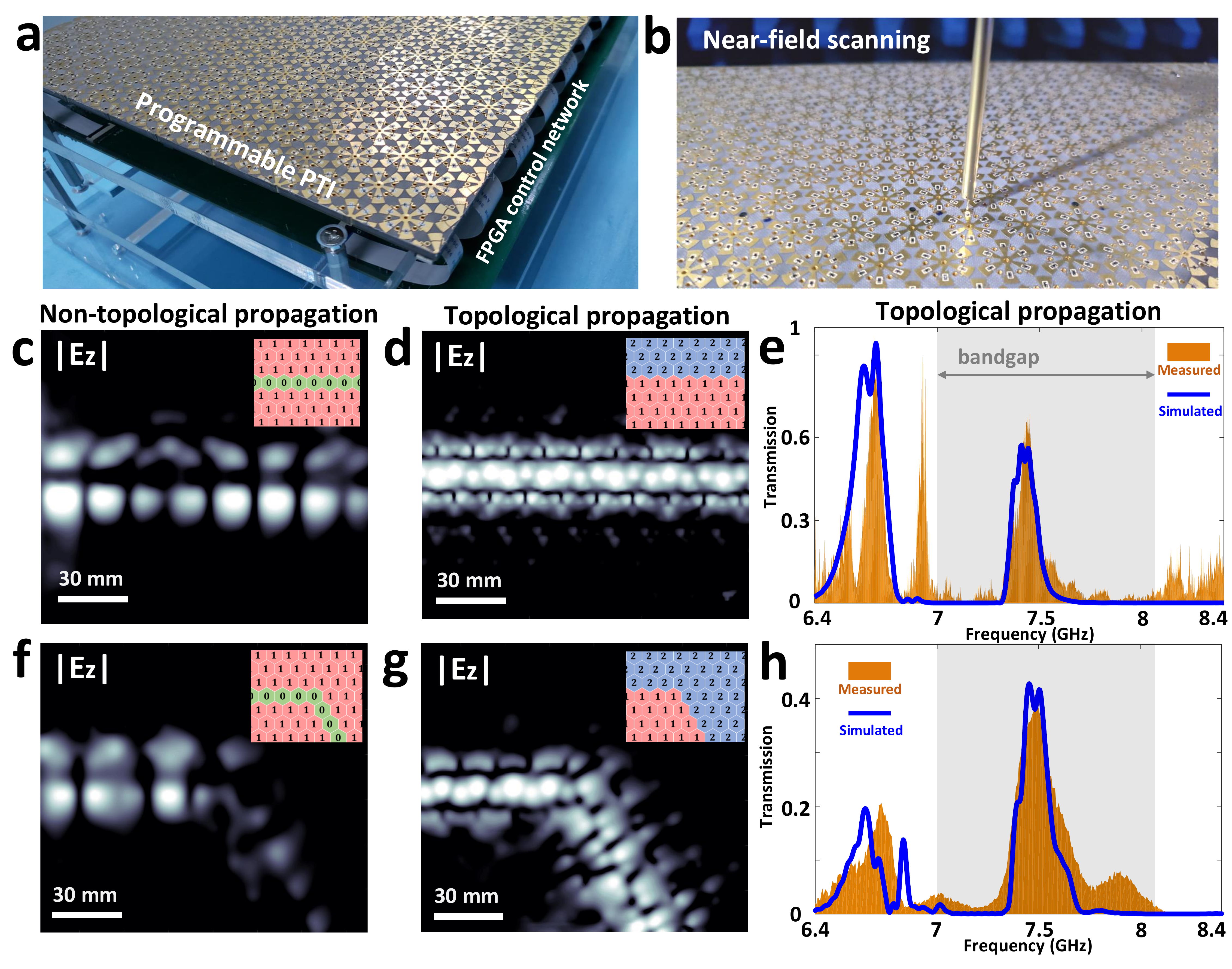}
\caption{\textbf{Experimental results of reprogrammable non-topological and topological propagation routes.} \textbf{a}, Photograph of the fabricated RPTI. \textbf{b}, Experimental setup of near-field scanning. \textbf{c}, \textbf{d}, Measured near-field distributions of $|E_z|$ in non-topological and topological straight waveguides at $7.45$ GHz, respectively. The corresponding coding patterns are depicted in insets. \textbf{e}, Comparison of measured and simulated transmissions of a topological straight waveguide. \textbf{f}, \textbf{g}, Measured near-field distributions of $|E_z|$ in non-topological and topological 60-degree-bend waveguides at $7.45$ GHz. \textbf{h}, Comparison of measured and simulated transmissions of a topological 60-degree-bend waveguide.}
\label{fig:TopoCompar}
\end{figure}

In addition to the unidirectional propagation, the feature of immunity against sharp bends is also experimentally studied here. As illustrated in Fig. \ref{fig:TopoCompar}a, a fabricated RPTI is dynamically controlled using an FPGA control network. To verify the backscattering-immune propagation against sharp bends, different topological and non-topological propagation routes are programmed in our RPTI by properly encoding each of its unit cells via the FPGA controller, and the corresponding near-field distributions of $|E_z|$ are measured by a SNMM experimental setup (see Fig. \ref{fig:TopoCompar}b).

To be more specific, a straight waveguide propagation route is firstly studied in both non-topological and topological cases with the measured near-field distributions of $|E_z|$ at 7.45 GHz presented in Figs. \ref{fig:TopoCompar}c and \ref{fig:TopoCompar}d, respectively. The corresponding coding patterns used to implement the two propagation routes are indicated in insets. These two measured results show the microwave in both cases can propagate very well. In Fig. \ref{fig:TopoCompar}e, the simulated and measured transmissions of the straight topological propagation route are compared, and very good agreement is observed. We also show in Extended Data Fig. 7a that the topological propagation route exhibits broadband transmission within the bandgap, and its maximum transmission is more than $3$ times larger than that of the non-topological propagation route. In addition to the straight propagation route, we further study the case of a 60-degree-bend propagation route. As shown in Figs. \ref{fig:TopoCompar}f and \ref{fig:TopoCompar}g, a strong backscattering induced by the bend is observed in the non-topological case, whereas in the topological case the propagation remains immune to the bend. The corresponding transmission spectra are compared in Fig. \ref{fig:TopoCompar}h and Extended Data Fig. 7b, where one can see that the transmission of the topological 60-degree-bend propagation route is slightly reduced as compared to the topological straight route, whereas the transmission of the non-topological 60-degree-bend propagation route is so small that its transmitted energy can be neglected. These results prove that the transmission characteristics of bends in the topological case are much superior to those in the non-topological case.

\section*{Ultrafast multi-channel optical analog-digital converter}

\begin{figure}[t!]
\centering
\includegraphics[width=15 cm]{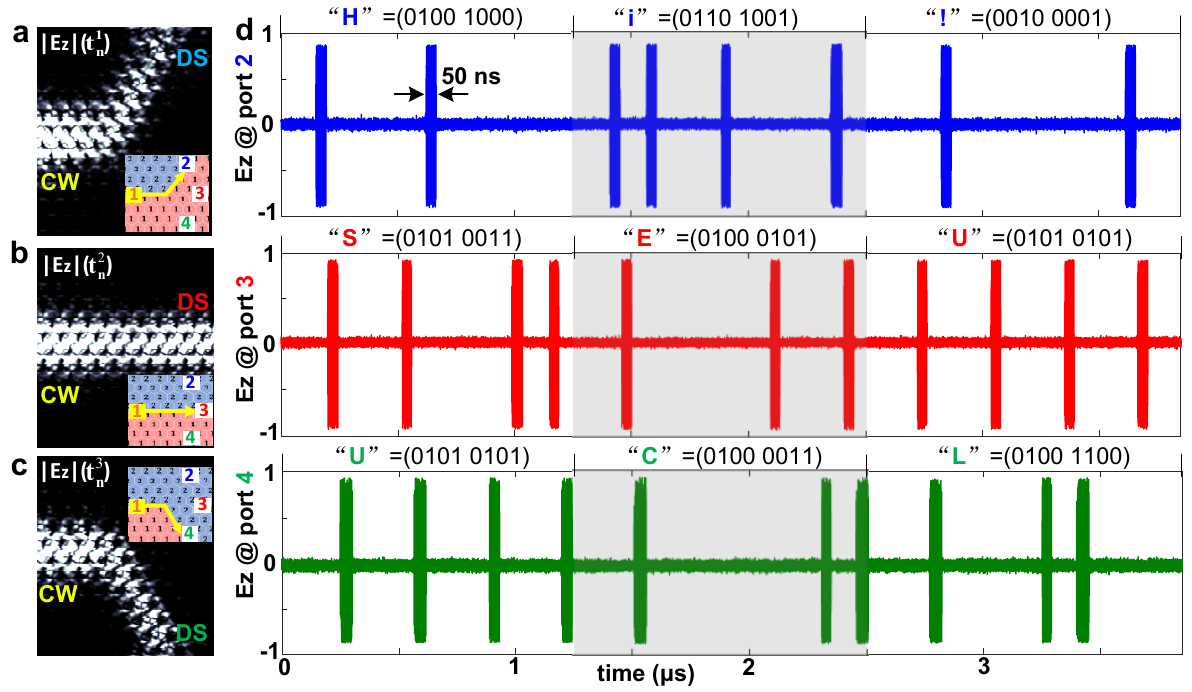}
\caption{\textbf{Proof-of-principle experiment of a reprogrammable multi-channel optical analog-digital converter.} \textbf{a}-\textbf{c}, Encoded propagation routes and the corresponding simulated field distributions of $|E_z|$ as a function of time $t_n$. Here, $t_n^1=t_0+3n\Delta t$, $t_n^2=t_0+(3n+1)\Delta t$, and $t_n^3=t_0+(3n+2)\Delta t$, where $n=0,1,2,3, ... $, $\Delta t$ is the switching period, and $t_0$ is the start time of the coding. A continuous wave (CW) is pumped at the input port 1. Because of the high-speed switching of topological propagation routes, the input CW is modulated directly as a digital signal (DS) detected at the output port. \textbf{d}, Measured output signal of each port. Three ASCII sequences (i.e. “Hi!”, “SEU”,“UCL”) are modulated and transmitted in our RPTI, and the switching time of topological propagation route can be as short as $50$ ns.}
\label{fig:OptADC}
\end{figure}

Due to the flexible programmability, many existing photonic topological functionalities can be readily implemented in our universal RPTI platform. More importantly, our RPTI can also be used to achieve some unprecedentedly dynamic functionalities that cannot be implemented in static PTIs or low-speed reconfigurable PTIs. Here, a proof-of-principle experiment of a topologically protected multi-channel optical analog-digital converter with ultrafast control is demonstrated in Fig. \ref{fig:OptADC}. To simplify our demonstration, four waveguide ports are designed in our RPTI as illustrated in Figs. \ref{fig:OptADC}a - \ref{fig:OptADC}c, and a continuous wave (CW) with frequency $7.45$ GHz is pumped at the input port $1$. By encoding different topological propagation routes, we can dynamically switch the output port from ports $2$ to port 3 or port $4$, thus the input CW analog signal can be directly discretized into different digital signals (DS) at the output ports. To demonstrate this remarkable feature, the input CW signal is modulated as three ASCII sequences (the corresponding glyph sequences are “Hi!”, “SEU” and “UCL”), and the three output digital signals are measured experimentally, as shown in Fig. \ref{fig:OptADC}d. The measured results show that the switching time in this system is as short as $50$ ns. To the best of our knowledge, this is the fastest switching speed of topological propagation routes, namely more than seven orders of magnitude faster than that of previous work \cite{che16NatMat}. Moreover, since only one topological propagation channel exists at each time step, the interference among the output ports is negligible, even when the output ports are closed to each other. This low cross-talk feature is crucial in practical high-fidelity digital communications.

To summarize, we have theoretically proposed and experimentally demonstrated a reprogrammable plasmonic topological insulator with ultrafast switching speeds at nanosecond scale. By investigating the band diagrams of the proposed RPTI, we shown that a topological bandgap emulating QVHE can be created by digitally encoding its unit cells via switching electrically the PIN diodes “on” and “off”. As each unit cell can be independently programmed, it was experimentally demonstrated that valley-Hall topological light propagation routes in our RPTI can be dynamically reconfigured. Moreover, we experimentally implemented a topologically protected multi-channel optical analog-digital converter to illustrate the ultrafast control feature of our RPTI, and demonstrated that the switching time in our case is as short as $50$ ns, which is more than seven orders of magnitude faster than that of previous work. Our work on RPTI holds great promise for the development of intelligent and agile topological photoelectric devices and high-speed reconfigurable digital communication systems.

\section*{Methods}

\textbf{Numerical simulations}. The numerical simulations were performed using the finite-element method based software COMSOL Multiphyiscs. In Fig. \ref{fig:PhysSche}b, the metallic material (yellow) is copper and the dielectric substrate (semitransparent grey) is a widely-used commercial PCB board (F4B), whose relative permittivity is $2.65$ and loss tangent is $0.001$. The PIN diodes (bold pink) are modelled by lumped elements. The “off” state is modelled by an inductor-capacitor series circuit, where the inductance $L=0.4$ nH and the capacitance $C=40$ fF. The band diagrams in Fig. \ref{fig:PhysSche}d were evaluated at the outer-arm length $s = 1.7$ mm. The “on” state is modelled by a resistor-inductor series circuit, where the resistance $R=2.2~\Omega$ and the inductance $L=0.32$ nH. The projected band diagrams in Figs. \ref{fig:TopoInterf}a and \ref{fig:TopoInterf}b were evaluated using a supercell containing 10 unit cells on either side of the domain wall, and periodic boundary conditions were imposed at left and right sides.

\noindent \textbf{Sample fabrication}. Our RPTI sample was fabricated in multiple-layer lamination process of PCB technique. In the layer stack management, as presented in Extended Data Fig. 1b, $L_1$ and $L_2$ are printed on the top and bottom facets of a F4B substrate (3 mm thickness), whereas $L_3$ and bias-voltage-line layers (Extended Data Fig. 6b) are printed on the top and bottom facets of a FR4 substrate (0.5 mm thickness), whose relative permittivity is 4.3 and loss tangent is 0.025. A prepreg layer (FR4 with 0.2 mm thickness) for bonding the F4B and FR4 substrates is used between $L_2$ and $L_3$ layers. The thickness of the printed copper film in each layer is $35~\mu$m. The overall size of our RPTI is about $465\times221$ mm$^2$, containing 488 digital units with 2928 PIN diodes. Commercially available Aluminum Gallium Arsenide PIN diodes (MADP-000907-14020) are employed in the fabrication.

\noindent \textbf{Experimental measurements}. A scanning near-field microwave microscopy is employed to measure the near electric fields and transmissions. Our SNMM is composed of vector network analyser (Agilent N5230C), low-noise amplifier, and several phase-stable coaxial cables. Two coaxial probes that act as the emitter and receiver are employed in measurements. One probe is soldered on the edge of the RPTI to generate a plasmonic mode, whereas the other one is fixed on a scanning support to scan the near-field distributions. During the measurements, the RPTI is placed on an acrylic supporter, as demonstrated in Extended Data Fig. 6a. Since each digital unit requires two voltage-control signals to realize the 2-bit states (Extended Data Fig. 1), our RPTI has 976 independent channels for voltage control. In order to dynamically steer the coding pattern of our RPTI, we use three field programmable gate arrays with a total of 1200 digital-voltage channels. The PIN diodes are connected to FPGA through white flexible printed circuit (FPC) cables, as shown in Extended Data Fig. 6c. In the measurement of the multi-channel optical analog-digital converter, a high-frequency oscilloscope (Agilent DSO9013A) and a signal generator (Agilent E8267D) are used to measure the wave forms in the time domain.


\subsection*{Acknowledgements}
This work was supported by European Research Council (ERC), Grant Agreement no. ERC-2014-CoG-648328; National Key Research and Development Program of China (Grant nos. 2017YFA0700201, 2017YFA0700202, 2017YFA0700203); and the 111 Project (Grant no. 111-2-05).

\subsection*{Author contributions}
JWY, NCP and TJC conceived the idea, JWY performed the theoretical analysis and numerical simulations, ZL helped with the theoretical analysis, QM and QX performed the experiments. NCP and TJC supervised the project. All authors contributed to the manuscript preparation.

\subsection*{Competing interests}
The authors declare no competing interests.

\subsection*{Additional information}
Correspondence and requests for materials should be addressed to N. C. Panoiu and T. J. Cui.

\end{document}